\documentstyle[12pt]{article}
\newcommand{\drawsquare}[2]{\hbox{%
\rule{#2pt}{#1pt}\hskip-#2pt
\rule{#1pt}{#2pt}\hskip-#1pt
\rule[#1pt]{#1pt}{#2pt}}\rule[#1pt]{#2pt}{#2pt}\hskip-#2pt
\rule{#2pt}{#1pt}}

\newcommand{\PSbox}[3]{\mbox{\rule{0in}{#3}\includegraphics{#1}\hspace{#2}}}
\newcommand{\Yfund}{\raisebox{-.5pt}{\drawsquare{6.5}{0.4}}}

\newcommand{\jref}[4]{{\it #1} {\bf #2}, #3 (#4)}

\newcommand{\NPB}[3]{\jref{Nucl.\ Phys.}{B#1}{#2}{#3}}

\newcommand{\PLB}[3]{\jref{Phys.\ Lett.}{#1B}{#2}{#3}}

\newcommand{\PRD}[3]{\jref{Phys.\ Rev.}{D#1}{#2}{#3}}

\begin{document}
\begin{titlepage}
\begin{center}
{\hbox to\hsize{hep-th/9708082 \hfill  LBNL-40695}}

\bigskip

\bigskip
\vspace{3\baselineskip}

{\Large \bf Duality in $Sp$ and $SO$ Gauge Groups from M Theory }

\bigskip

\bigskip

{\bf Csaba Cs\'aki\footnote{Research Fellow, Miller Institute for 
Basic Research in Science, 2536 Channing Way, Berkeley, CA 94720} }\\
\bigskip
{\small \it Theory Group, Lawrence Berkeley National Laboratory, \\ 
Mail Stop 50A-5101, Berkeley, CA 94720. \\}
\bigskip
{\bf and} \\
\bigskip
{\bf Witold Skiba\footnote{Address after September 1: Department of Physics, 
University of California at San Diego, La Jolla,  CA 92093.} }\\

\bigskip
{ \small \it Center for Theoretical Physics,

Massachusetts Institute of Technology, Cambridge, MA 02139.}

\bigskip

{\tt csaki@mit.edu, skiba@mit.edu}

\bigskip

\vspace*{1cm}
{\bf Abstract}\\
\end{center}

\noindent
We describe fivebrane configurations in M theory whose 4-d spacetime
contains $N=1$ supersymmetric $Sp$ or $SO$ gauge fields and fundamentals
of these groups. We show how field-theory dualities for $Sp$ and $SO$
groups can be derived using these fivebrane configurations  in M theory.
  
\bigskip

\bigskip

\end{titlepage}

\section{Introduction}
It is difficult to understand the dynamics behind the electric-magnetic
duality in supersymmetric field theory~\cite{Seiberg} in the context of
field theory. String theory brane dynamics~\cite{Polchinsky} turns out
to provide new information about the phenomena of duality in supersymmetric
field theories. Here, we will explore the M theory approach for learning
about the low-energy limit of $N=1$ supersymmetric field theories~[3-9].
So far, many field-theory results have been rederived
using various configurations of branes in string theory~[10-19].
Depending on the geometry of the brane setup one obtains different gauge
theories with varying amount of supersymmetry on the world-volume of branes.
In the given examples continuous deformations of branes lead to theories 
with the same infrared
dynamics, while the high-energy field content can change under deformations.
Many examples of duality have been confirmed this way. Similar insights
have been gained using F theory approach~[20-27].

However, these brane configurations  in the context of string theory
involve a singularity at the points where the branes join. To avoid
these type of singularities it has been suggested to consider the
given brane setup embedded into M theory~\cite{Lerche,Witten1}. 
The advantage of M theory
is that it smoothes out many of the singularities encountered at 
the joining of branes. For example, D-4 branes can be thought of as
5-branes wrapped around the eleventh dimension. 
This way, the brane setup corresponding to interesting 4-d field 
theories will
from a single 5-brane surface in M theory. 
For example, a common setup in type IIA string
theory for studies of dualities is a series of parallel NS 5-branes
connected by Dirichlet 4-branes. In M theory, D-4 branes become 5-branes
wrapped around compact eleventh dimension. Therefore branes connect smoothly
in such a setup. Using single brane configuration avoids singularities present
in other approaches. 
With this approach, Witten
computed 
the elliptic curves describing the Coulomb
branch of $N=2$ $SU(N)$ theory using M theory in 
Ref.~\cite{Witten1}. Subsequently, the authors of Ref.~\cite{SpSO1,SpSO2,noyy}
generalized these results to other classical groups. $N=1$ theories can be
studied on the world-volume of branes in M theory as well. Several authors
obtained results about the confining phase of $N=1$
supersymmetric QCD, dynamically
generated superpotentials and also gaugino condensation~[3-7, 9]. Recently, 
Schmaltz and Sundrum have pointed out that the embedding of the type IIA
theory can also avoid the singularities that one encounters in string
theory when moving branes across each other. They have shown 
how to derive 
electric-magnetic duality in $N=1$ $SU(N)$ theories with fundamentals
from M theory~\cite{MartinRaman}. In their setup a single 
M theory 5-brane describes both the electric and the 
magnetic theories of Seiberg's SUSY QCD, if non-vanishing masses for the
quarks are assumed. This way they obtained a smooth interpolation
between the electric and the magnetic descriptions of
SUSY QCD. However, a non-vanishing mass term for this smooth
interpolation was crucial, the M theory surface for the massless case is still
singular. 

In this paper
we generalize the results of Ref.~\cite{MartinRaman} and derive duality
in $Sp$ and $SO$ gauge groups~\cite{IntSeib,IntPoul}. We first describe
brane configurations in type IIA string theory, which we later interpret
in the context of M theory. We will show that the same curve describes the 
original $Sp(2N)$ $[SO(N)]$ theories as well as the dual
$Sp(2F-2N-4)$ $[SO(F-N+4)]$ theories. For the $Sp$ theories we also
extend our results to a setup with finite fourbranes, where the
antisymmetric meson field of the dual theory will emerge explicitly.

\section{Semi-infinite Brane Configuration for $Sp(2N)$}

We begin by considering brane configurations in type IIA string theory.
We denote spacetime coordinates by $x^0,x^1,\ldots,x^9$, where
$x^0,\ldots,x^3$ denote the usual 4-d spacetime. For future reference let
us define $v=x^4+i x^5$ and $w=x^7+i x^8$. Once we move on to M theory we 
will denote the eleventh dimension by $x^{10}$. The $x^{10}$ coordinate
is periodic under $x^{10}\rightarrow x^{10}+R$, where $R$ is the 
compactification radius. From the string theory point of view $R=g_s$,
so small radius limit is equivalent to weakly-coupled string theory.
It will be later useful to define $s=x^6+ix^{10}$,
$t=\exp (-\frac{s}{R})$. Also for convenience, we choose the units
such that the string scale is set to one, $m_s=1$. 

\begin{figure}[!ht]
  \PSbox{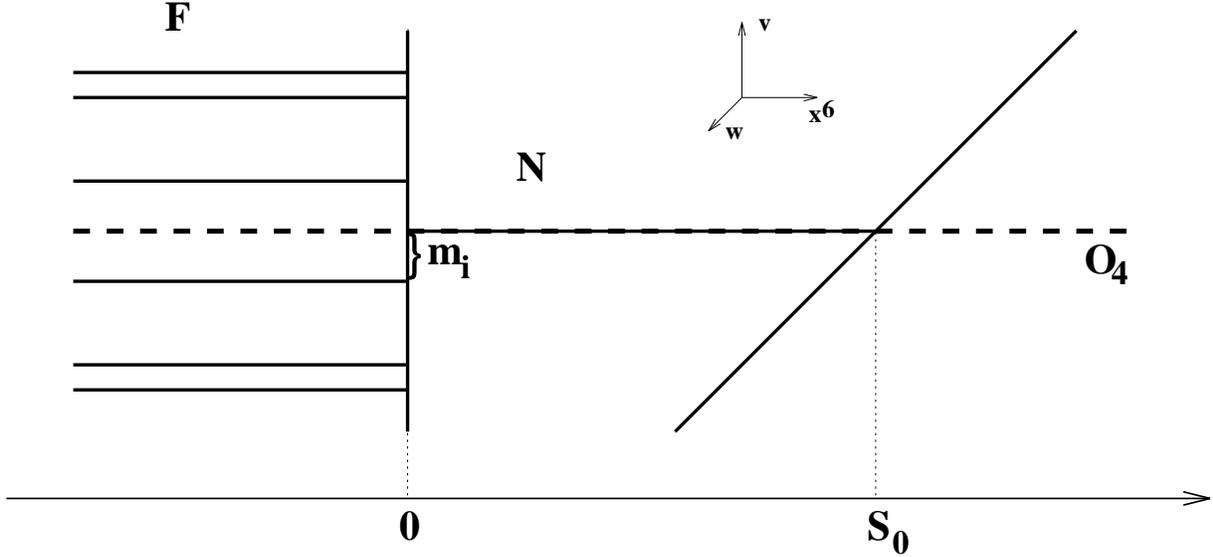 hscale=70 vscale=70 hoffset=0 
         voffset=0}{13.7cm}{7.5cm} 
  \caption{Brane configuration of the electric theory.}
  \label{fig:1}
\end{figure}

We use a brane configuration similar to that of Ref.~\cite{MartinRaman}.
We consider Dirichlet 4-branes stretched between NS 5-branes. We will take
some 4-branes to be semi-infinite. Our configuration is illustrated in
Fig.~\ref{fig:1}. All branes fill the 4-d spacetime and are placed at $x^9=0$.
There are two 5-branes: one at $x^6=0$ occupying $x^4$ and $x^5$ ($5_v$),
and another one at $x^6=s_0$ occupying $x^7$ and $x^8$ ($5_w$).
There are $N$ 4-branes suspended between the 5-branes and also $F$ 4-branes
extending from the $5_v$ brane to minus infinity in the $x^6$ direction.
In order to obtain $Sp$ or $SO$ gauge groups we need to make an orientifold
projection~\cite{orientifold}. Orientifold projection combines
spacetime symmetry and a parity inversion on the world-sheet.
Under this projection the spacetime coordinates transform as 
\begin{displaymath}
(x^4,x^5,x^7,x^8,x^9) \longrightarrow (-x^4,-x^5,-x^7,-x^8,-x^9).
\end{displaymath}
Modding out the spacetime by this transformation is indicated in 
Fig.~\ref{fig:1} by an orientifold 4-plane. The orientifold 4-plane extends in
$x^1$, $x^2$, $x^3$ and $x^6$ directions. 4- and 5-branes are placed
symmetrically with respect to the orientifold. The parity projection,
$\Omega$, allows for $\Omega^2=\pm 1$. The 4-d gauge group is symplectic
when $\Omega^2=-1$, and it is orthogonal otherwise. 
In order to generate non-vanishing masses for the flavors, the 
semi-infinite 4-branes are assumed to be 
non-overlapping. The distances of these 4-branes
from the orientifold plane correspond to masses of fundamental fields $m_i$.
We will assume that all these masses are different. Strings connecting the
finite 4-branes correspond to massless vector fields of $Sp$ ($SO$) gauge
groups. The separation between the $5_v$ and $5_w$ branes in the $x^6$
direction is related to the gauge coupling of the $4-d$ theory:
\begin{displaymath}
  \frac{8 \pi^2}{g_4^2} \sim \frac{s_0}{g_s}. 
\end{displaymath}

Having described the topology of the brane configuration we lift the above
setup into M theory, which also allows us to take the limit of large
string coupling. 
The 4-branes gain an extra
dimension, since they are compactified on $x^{10}$ in the M theory 
description, and connect smoothly with
the $5_v$ and $5_w$ branes. We first consider the case of $Sp$ gauge group
and determine the curve corresponding to the M theory
configuration~\cite{Witten1,MartinRaman}. The setup of Fig.~\ref{fig:1} will
correspond to an $Sp(2N)$ gauge theory with $2F$ fundamental fields
which have a mass term, $m QQ$, in the superpotential. In M theory
this setup will occupy $R^4\times \Sigma$, where $\Sigma$ is a one
complex dimensional Riemann surface. In the case of $Sp(2N)$ this
surface is described by the curve
\begin{eqnarray}
 t \prod_{i=1}^{F} (v^2-m_i^2) & = & \xi v^{2N+2}, \nonumber \\
 v w & = & \zeta.
\end{eqnarray}

It is helpful to identify the symmetries associated with rotations in the
$v$ and $w$ planes. These symmetries are anomalous so the scale of 4-d
theory, $\Lambda_{Sp}$, transforms under these symmetries. The mass
terms, $m Q_i Q_j$, will be kept invariant under the transformations by
assigning appropriate charges to $m$. The mesons, $M$, in the dual theory
which we will describe later, always carry the charges of quark bilinears.
The table of charges is as follows:
\begin{displaymath}
  \begin{array}{c|ccccc}
       & Q & m & \Lambda_{Sp}^{3 (N+1)-F} & v & w \\ \hline
   R_v & 0 & 2 & 2N+2-2F                  & 2 & 0 \\
   R_w & 1 & 0 & 2N+2                     & 0 & 2
  \end{array}
\end{displaymath}
Using these symmetries we can identify the parameters $\xi$ and $\zeta$:
\begin{eqnarray}
 t \prod_{i=1}^{F} (v^2-m_i^2) & = & v^{2N+2}
     \left({\rm Pf} m \right)^{2\frac{F-N-1}{F}}, \nonumber \\
 v w & = & \Lambda_{Sp}^{\frac{3 (N+1)-F}{N+1}}
     \left({\rm Pf} m \right)^{\frac{1}{N+1}} .
\end{eqnarray}

There are several consistency checks one can perform on the above form
of the curve. We will check how the curve behaves for large values of
$v$ and $w$, in which case we expect to reproduce the perturbative
limit of the gauge theory. The symmetries of the curve in such limits
should correspond to symmetries of the microscopic gauge theory.
First, we can examine the asymptotic behavior of the curve for large
$v$:
\begin{equation}
\label{asympt1}
  t v^{2F-2N-2} \sim \left({\rm Pf} m\right)^{2\frac{F-N-1}{F}}, \; \;
  w \sim 0,
\end{equation}
which is indeed symmetric under $R_v$ and $R_w$. For large $w$, the
curve can be approximated as
\begin{equation}
\label{asympt2}
  t w^{2N+2} \sim \Lambda_{Sp}^{2 [3(N+1)-F]}
    \left({\rm Pf} m\right)^{2\frac{F-N-1}{F}}, \; \;
  v \sim 0.
\end{equation}
Here, one needs to take into account transformation properties of
$\Lambda_{Sp}$ under anomalous symmetries $R_v$ and $R_w$. From field
theory one expects the presence of non-anomalous discrete symmetries.
In either limit, $R_v$ rotation by $\frac{\pi}{2 N +2-2F}$ and $R_w$
rotation by $\frac{\pi}{2N+2}$ leave the curve invariant. 
$R_v$ is explicitly broken by the mass term $m$, while $R_w$ is broken
to its $Z_{4(N+1)}$ subgroup. 
As in Ref.~\cite{MartinRaman} we can calculate the separation of the
$5_v$ and $5_w$ branes in the $s$ direction. This can be done by going to large
values of $v$ on the $5_v$ brane and to large values of $w$ on the
$5_w$ brane. Then using the above equations for $t$ we obtain that 
\begin{equation} 
\label{bending}
e^{-s_0/R}=\Lambda_{Sp}^{2[3(N+1)-F]},
\end{equation}
where $s_0$ is the distance between the two branes.
This equation just expresses the logarithmic bending of branes at large
distances. 
Note, that just like in the
$N=2$ case explained in Ref.~\cite{SpSO1,SpSO2}, we get a factor of 
two in front of the beta function in Eq.~\ref{bending}, which
corresponds to a rescaling of the gauge coupling constant 
and appears due to the non-conventional embedding of $Sp(2N)$ into
$SU(2N)$~\cite{SpSO1,SpSO2}. 

Let us now express the curve in terms of $t$ and $w$:
\begin{displaymath}
t (-1)^F \prod_{i=1}^{F} (w^2-w_i^2) =  w^{2F-2N-2}
     \left( \prod_{i=1}^{F} w_i \right)^{2\frac{N+1}{F}},
\end{displaymath}
where $w_i \propto \left( {\rm Pf} m \right)^\frac{1}{N+1}
    \Lambda^\frac{3 (N+1) -F}{N+1} \left(\frac{1}{m} \right)_i$.
This expression for $w_i$ 
looks exactly like the field-theory relation for the VEVs of the
meson operators, $M$, in the presence of mass terms $m Q Q$~\cite{IntPoul}:
\begin{displaymath}
  \langle  M_{ij} \rangle =
    \left[ 2^{N-1} \left( {\rm Pf} m \right) \right]^\frac{1}{N+1}
    \Lambda^\frac{3 (N+1) -F}{N+1} \left(\frac{1}{m} \right)_{ij}.
\end{displaymath}
After reabsorbing suitable numerical factors into the definition of $t$, 
the curve written in terms of $t$ and $w$, where we also replaced $w_i$ by
$\langle M \rangle_i$
\begin{eqnarray}
 t \prod_{i=1}^{F} (w^2-\langle M \rangle_i^2) & = & w^{2\tilde{N}+2}
     \left({\rm Pf} \langle M \rangle \right)^{2\frac{F-\tilde{N}-1}{F}},
                                          \nonumber \\
 v w & = & \tilde{\Lambda}_{Sp}^{\frac{3 (\tilde{N}+1)-F}{\tilde{N}+1} }
     \left({\rm Pf} \langle M \rangle \right)^{\frac{1}{\tilde{N}+1}} ,
\end{eqnarray}
where $\tilde{N}=F-N-2$, while $\tilde{\Lambda}_{Sp}$ is the scale of the
dual gauge group, and the scale of the magnetic theory is related to
the scale of the electric theory as $\tilde{\Lambda}_{Sp}^{3 (\tilde{N}+1)-F} 
\Lambda_{Sp}^{3 (N+1)-F}=16 (-1)^{F-N-1}\mu^F$. The scale $\mu$ is arbitrary
in field theory, however in this derivation $\mu$ is proportional to the
string mass scale. 

\begin{figure}[!ht]
  \PSbox{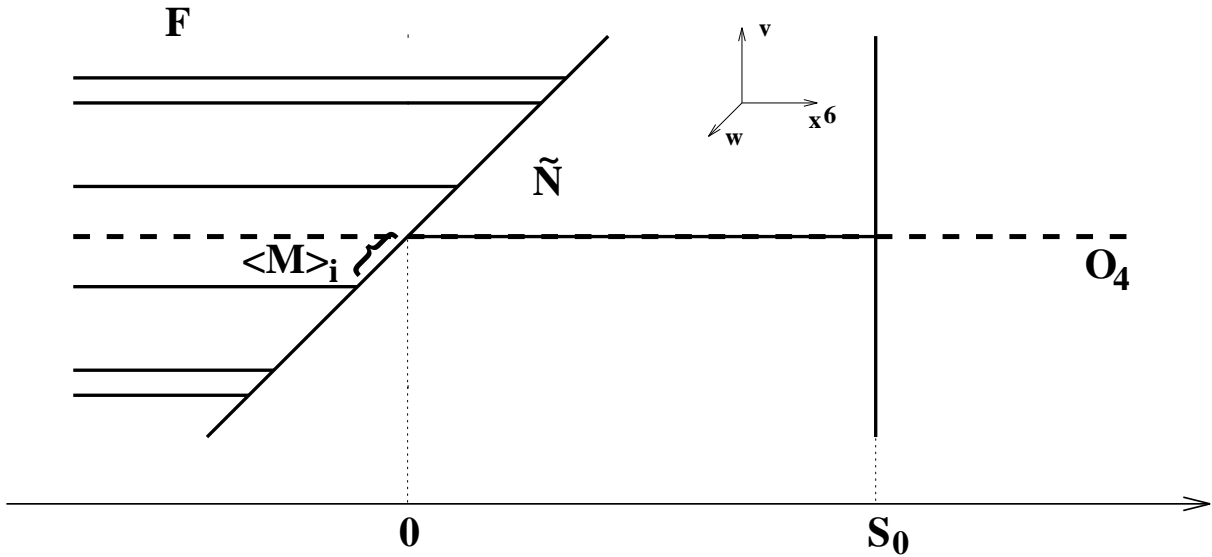 hscale=70 vscale=70 hoffset=0 
         voffset=0}{13.7cm}{7.5cm} 
  \caption{Brane configuration of the magnetic theory.}
  \label{fig:2}
\end{figure}

This curve looks exactly like the original one, except that it corresponds
to the dual gauge group $Sp(2 F - 2N -4)$, whose brane configuration is
illustrated in Fig.~\ref{fig:2}. The distances between 4-branes and the 
orientifold plane can now be identified with the expectation values
of the mesons in the magnetic theory.  By comparing with the original curve,
we can see that the expectation values of the meson fields play now the role
of the dual quark masses, which is exactly what we expect from field
theory~\cite{IntPoul}. 
Taking the limit $\tilde{\Lambda}_{Sp}\to 0$ (which corresponds to
taking $\Lambda_{Sp}\to \infty$) and then $R\to 0$ while keeping the
meson VEV fixed will result in the brane configuration displayed in 
Fig.~\ref{fig:2}. Thus we can
see the emergence of the dual $Sp(2F-2N-4)$ gauge group by looking
at the same M theory fivebrane in two different limits.

\section{Semi-infinite Brane Configuration for $SO(2N)$ and SO(2N+1)}

We now repeat the above analysis for $SO(N)$ groups. There are two cases
that need to be dealt with separately: when $N$ is even and $N$ is odd.
The brane configuration we use is that of Fig.~\ref{fig:1}. We first
consider $SO(2N)$ groups with $2F$ vectors. The charge assignment
under $R_v$ and $R_w$ rotation is the following:
\begin{displaymath}
  \begin{array}{c|ccccc}
       & Q & m & \Lambda_{SO}^{3 (2N-2)-2F} & v & w \\ \hline
   R_v & 0 & 2 & 2(2N-2)-4F                   & 2 & 0 \\
   R_w & 1 & 0 & 2(2N-2)                      & 0 & 2
  \end{array}
\end{displaymath}
We obtain the following curve in this case
\begin{eqnarray}
 t v^2 \prod_{i=1}^{F} (v^2-m_i^2) & = & v^{2N}
     \left( \det m \right)^{\frac{F-N+1}{F}}, \nonumber \\
 v w & = & \Lambda_{SO}^{\frac{3 (2N-2)-2F}{2N-2}}
     \left( \det m \right)^{\frac{1}{2N-2}} .
\end{eqnarray}
We can perform the same consistency checks we did in the case of $Sp$. In the
large $v$ limit we get
\begin{equation}
  t v^{2F-2N+2} \sim \left( \det m\right)^{\frac{F-N+1}{F}}, \; \;
  w \sim 0,
\end{equation}
while in the large $w$ limit
\begin{equation}
  t w^{2N-2} \sim \Lambda_{SO}^{ [3(2N-2)-2F]}
    \left( \det m\right)^{\frac{F-N+1}{F}}, \; \;
  v \sim 0.
\end{equation}
It is straightforward to check that the symmetries we expect from field
theory are properly reproduced. Here, the separation of $5_v$ and $5_w$ branes
in the $x^6$ direction gives due to brane bending
\begin{displaymath}
   e^{-s_0/R}=\Lambda_{SO}^{3(2N-2)-2F}.
\end{displaymath}

The meson fields in $SO(N)$ theories obtain the following VEVs when mass
terms for the quark fields are present
\begin{displaymath}
  \langle  M_{ij} \rangle =
    \left[ 16 \left( \det m \right) \right]^\frac{1}{2N-2}
    \Lambda^\frac{3 (2N-2) -2F}{2N-2} \left(\frac{1}{m} \right)_{ij}.
\end{displaymath}
Identical expression emerges when we express the curve in terms of $w$ and $t$,
which is a strong indication that the curve properly reproduces mesons
VEVs in the magnetic theory. We obtain the following form for the curve:
\begin{eqnarray}
 t w^2 \prod_{i=1}^{F} (w^2-\langle M_i \rangle^2) & = & w^{2\tilde{N}}
     \left( \det \langle M \rangle  \right)^{\frac{F-\tilde{N}+1}{F}},
    \nonumber \\
 v w & = & \tilde{\Lambda}_{SO}^{\frac{3 (2\tilde{N}-2)-2F}{2\tilde{N}-2)}}
 \left( \det \langle M \rangle  \right)^{\frac{1}{2\tilde{N}-2}},
\end{eqnarray}
where $\tilde{N}=F-N+2$, while
$\Lambda_{SO}^{3 (2 N-2) -2F} \tilde{\Lambda}_{SO}^{3 (2 \tilde{N} -2) -2F} =
  \mu^{2 F}$. The fundamental result is again confirmed.
The dual gauge group is $SO(2F-2N+4)$, which one obtains in the
$\tilde{\Lambda}_{SO}\to 0, R\to 0$ limit while keeping the
meson VEV fixed.

Let us now briefly summarize the same derivation for $SO(2N+1)$. As before, we
begin with the table of charge assignments.  
\begin{displaymath}
  \begin{array}{c|ccccc}
       & Q & m & \Lambda_{SO}^{3 (2N-1)-2F} & v & w \\ \hline
   R_v & 0 & 2 & 2(2N-1)-4F                   & 2 & 0 \\
   R_w & 1 & 0 & 2(2N-1)                      & 0 & 2
  \end{array}
\end{displaymath}
The curve describing the brane configuration for $SO(2N+1)$ is 
\begin{eqnarray}
 t v \prod_{i=1}^{F} (v^2-m_i^2) & = & v^{2N}
     \left( \det m \right)^{ \frac{2F-2N+1}{2F}}, \nonumber \\
 v w & = & \Lambda_{SO}^{\frac{3 (2N-1)-2F}{2N-1}}
     \left( \det m \right)^{\frac{1}{2 N-1}} .
\end{eqnarray}
We have checked the perturbative limits
\begin{eqnarray}
 v \rightarrow \infty ,&
 t v^{2F-2N+1} \sim \left( \det m\right)^{\frac{2F-2N+1}{2F}}, \; \;
  & w \sim 0, \nonumber \\
w \rightarrow \infty , &  t w^{2N-1} \sim \Lambda_{SO}^{ [3 (2N-1)-2F]}
    \left( \det m\right)^{\frac{2F-2N+1}{2F}}, \; \;
  & v \sim 0, \nonumber
\end{eqnarray}
and they indeed have the correct symmetry properties. Expressing the 
curve in the $t-w$ variables yields:
\begin{eqnarray}
 t w \prod_{i=1}^{F} (w^2-\langle M_i \rangle^2) & = & w^{2\tilde{N}}
     \left( \det \langle M \rangle  \right)^{\frac{2F-2\tilde{N}+1}{2F}},
    \nonumber \\
 v w & = & \tilde{\Lambda}_{SO}^{\frac{3 (2\tilde{N}-1)-2F}{2\tilde{N}-1}}
 \left( \det \langle M \rangle  \right)^{\frac{1}{2\tilde{N}-1}}.
\end{eqnarray}
Again, duality is reproduced properly. In the above 
 considerations we did 
see the emergence of the expectation values of the
meson fields, however the meson field itself corresponds to five dimensional
excitations and their coupling to the four dimensional theory is very weak.
We need to consider finite fourbranes
instead of the semi-infinite ones to overcome this problem. This will
be considered in the next section for the case of $Sp$ theories.

Note, that in the case of $SO(N)$ groups we had to restrict ourself
to theories with even number of vectors, even though there is no
field theory reason to do so. The string theory reason behind this
is that we wanted to give a mass to every vector. In the orientifold 
construction however one vector is necessarily massless in the case 
of odd number of vectors, since one vector has to lie on the
orientifold plane.

\section{Finite Brane Configuration for $Sp(2N)$}

Above, we have considered the case where the fourbranes giving rise to the
fundamentals of the $SO$ and $Sp$ groups are semi-infinite. We saw in this
picture how the dual gauge groups are emerging from one and the same
M theory fivebrane configuration, but the massless meson fields required for
Seiberg's duality were missing. In order to get these fields, we consider
the finite brane construction analogous to the one presented in 
Ref.~\cite{MartinRaman}. We will consider only the $Sp(2N)$ case, while the
other groups can be worked out similarly. 

\begin{figure}[!ht]
  \PSbox{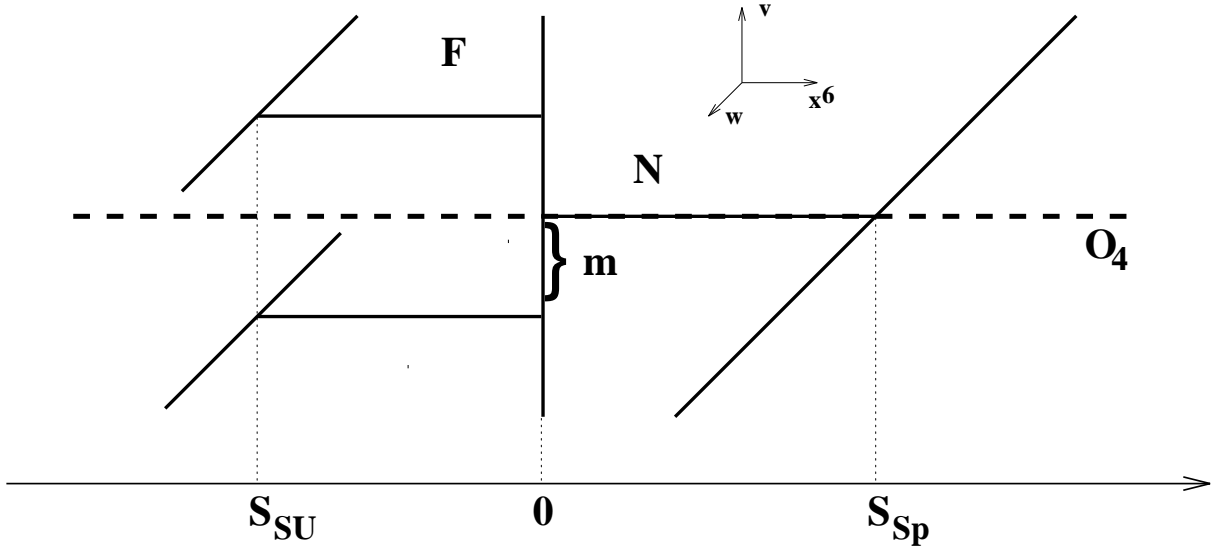 hscale=70 vscale=70 hoffset=0 
         voffset=0}{13.7cm}{7.5cm} 
  \caption{Finite brane configuration of the electric theory.}
  \label{fig:3}
\end{figure}

The construction of \cite{MartinRaman} for SUSY QCD started with a brane
setup similar to the one presented above but the semi-infinite branes 
terminated in additional fivebranes filling out the w-plane. Just like
in \cite{MartinRaman} we consider the case with a common mass 
$m$ for the $Sp$ fundamentals only. 
Since we are interested in having non-vanishing masses for every fundamental
of $Sp(2N)$
the $F$ fourbranes must be placed symmetrically above and below
the O4 orientifold. Because we are considering the case when these fourbranes
terminate on additional fivebranes, part of the flavor symmetry will
be gauged. The full flavor symmetry is $SU(2F)$, however it is
broken by the mass term $m$. Thus the gauged part of the flavor 
symmetry will be only $SU(F)$. 
This brane setup is depicted in Fig.~\ref{fig:3}, while the spurious symmetries
of the field theory are summarized in the table below.

\[
\begin{array}{c|cccc}
& SU(F) & Sp(2N) & R_v & R_w \\ \hline
Q & \Yfund & \Yfund & 0 & 1 \\
\bar{Q} & \overline{\Yfund} & \Yfund & 0 & 1  \\
\Lambda_{Sp}^{3(N+1)-F} & 1 & 1 & 2N+2-2F & 2N+2 \\
\Lambda_{SU}^{3F-2N} & 1 & 1 & 2F-4N & 2F \\
m & 1 & 1 & 2 & 0 \\
v & 1 & 1 & 2 & 0 \\
w & 1 & 1 & 0 & 2 \end{array} \]
There is a superpotential term $mQ\bar{Q}$  present in the field theory.
In field theory this superpotential results in gaugino condensation with
expectation value
\[ {\rm Tr} \langle Q\bar{Q} \rangle = (N+1)2^{\frac{N-1}{N+1}}
\left[ m^{F-N-1}\Lambda_{Sp}^{3(N+1)-F}\right]^{\frac{1}{N+1}}+
F\left[ m^{2N-F}\Lambda_{SU}^{3F-2N}\right]^{\frac{1}{F}}. \]
In M theory, the curve describing the $\Sigma$ Riemann surface 
for this brane setup is given by
\begin{eqnarray}
&& t(v^2-m^2)^F= v^{2N+2} m^{2F-2N-2} \nonumber \\
&&w= \frac{m}{v} \left[ m^{F-N-1}\Lambda_{Sp}^{3(N+1)-F}\right]^{\frac{1}{N+1}}
+ \frac{m^2}{v^2-m^2}\left[m^{2N-F} \Lambda_{SU}^{3F-2N}\right]^{\frac{1}{F}}.
\end{eqnarray}
One can check that this curve obeys the spurious $R_v$ and $R_w$ symmetries
as well as reproduces the correct classical limits:
\begin{eqnarray}
&& v\rightarrow m, \; t\rightarrow \infty ,\; w\rightarrow \infty, \;
t\sim w^{F}\Lambda_{SU}^{3F-2N} m^{F-2N}, \nonumber \\
&&  v\rightarrow 0, \; t\rightarrow 0 ,\; w\rightarrow \infty, \;
t\sim \frac{\Lambda_{Sp}^{2[3(N+1)-F]}m^{2F-2N-2}}{w^{2(N+1)}}, \nonumber \\
&&  v\rightarrow \infty , \; t\rightarrow 0 ,\; w\rightarrow 0, \;
t\sim \left( \frac{m}{v}\right)^{2F-2N-2}.
\end{eqnarray}

\begin{figure}[!ht]
  \PSbox{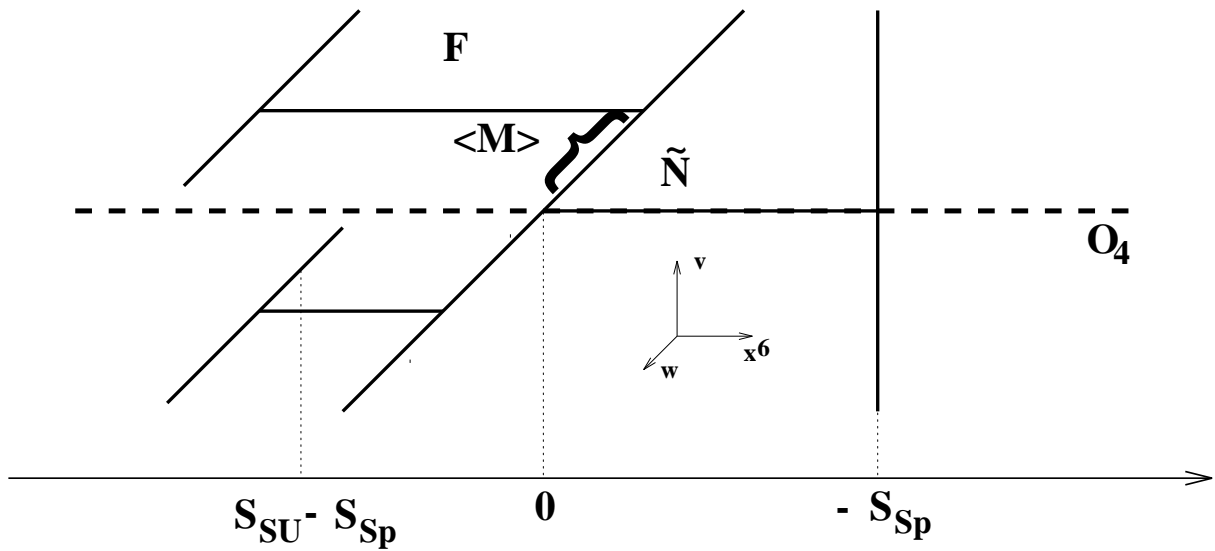 hscale=70 vscale=70 hoffset=0 
         voffset=0}{13.7cm}{7.5cm} 
  \caption{Finite brane configuration of the magnetic theory.}
  \label{fig:4}
\end{figure}

The weakly coupled string theory is reached by taking $R\rightarrow 0,
\Lambda_{SU},\Lambda_{Sp}\rightarrow 0$. Taking 
\begin{equation}
\Lambda_{Sp}^{3(N+1)-F}=e^{-S_{Sp}/R},\;  \Lambda_{SU}^{3F-2N}= 
e^{S_{SU}/R}, 
\end{equation}
where $S_{Sp}>0, S_{SU}<0$, and $m$ is fixed reproduces the 
setup of Fig.~\ref{fig:3}, with the fivebranes positioned at 
$x_6=S_{SU},0,S_{Sp}$. Duality can be obtained by taking the 
$\Lambda_{Sp}\gg 1$ limit, which will now imply $S_{Sp}<0$. Taking also
the $R\rightarrow 0$ limit while keeping $S_{Sp},S_{SU}$ and the
meson VEV fixed gives the brane setup depicted in Fig.~\ref{fig:4}.
The required  scaling of the mass is given by
\[ m^{F-N-1}\sim e^{S_{Sp}/R}. \]
Thus one can see that in the classical limit one also needs to take $m\to 0$.
We can read off the low-energy degrees of freedom as the string excitations
connecting the D4 branes in Fig.~\ref{fig:4}. The $Sp(2F-2N-4)$ gauge bosons 
arise from the strings between the color branes on the right in
Fig.~\ref{fig:4}, while the $F$ flavors of $Sp$ are the strings
connecting color and flavor branes.

The fivebranes on the left are now parallel and thus the flavor branes can 
slide freely along the $w$ direction, giving rise to the meson 
field $M$. To understand the properties of this meson field we note that
the left hand side of Fig.~\ref{fig:4} is $N=2$ supersymmetric, so there
must be an adjoint field of the $SU(F)$ gauge group  present. This is 
however only part of the antisymmetric meson field needed for the $Sp$-duality.
Note however, that there are additional (massive) four dimensional
excitations, which correspond to strings connecting the fourbranes above and
below the orientifold. These together with the adjoint of $SU(F)$ exactly
make up an antisymmetric tensor of $SU(2F)$, which then can be identified
with the meson field in the $Sp$ duality. 
Thus in this case there are no
missing components of the meson field, rather in the $m\neq 0$ some 
components of the meson field are massive, 
which is in agreement with the field theory
results.

\section*{Acknowledgments}
We thank Martin Schmaltz for several useful discussions and to Jan de Boer,
Kentaro Hori, Shamit Kachru and Hiroshi Ooguri for reading
the ma\-nus\-cript. This work has been supported in part by the Director, 
Office of Energy Research, Office of Basic Energy Sciences, of the US 
Department of Energy under Contract No. AC03-76SF00098 and in part by
the US Department of Energy under Cooperative Agreement No. DE-FC02-94ER40818.

\end{document}